\documentstyle[preprint,aps,epsfig]{revtex}
\begin{document}
\title{Geomagnetic Effects  on Atmospheric Neutrinos}

\author{Paolo Lipari$^2$, Todor Stanev$^1$ \& T.K. Gaisser$^1$\\
$^1$Bartol Research Institute\\
University of Delaware, Newark, DE 19716\\
$^2$ Dipt. di Fisica and INFN\\
Universit\`a di Roma I, Piazzale A. Moro 2\\
00185 Roma, Italy}
\maketitle
\begin{abstract}
Geomagnetic  effects  distort  the zenith angle  distribution
of  sub--GeV and few--GeV atmospheric  neutrinos, breaking 
the up--down  symmetry that would be present 
in the absence of  neutrino oscillations and without a geomagnetic field.
The  geomagnetic effects  also produce a  characteristic  azimuthal 
dependence of the $\nu$--fluxes,  related to the well known east--west
effect,  that should be detectable in neutrino experiments of sufficiently
large mass.  We discuss these effects quantitatively.
Because the azimuthal dependence is in first order 
independent of any oscillation effect, it is a useful
diagnostic  tool for studying possible
systematic effects in the search for neutrino  oscillations.
\end{abstract}

\section {Introduction}
The flux  of atmospheric  neutrinos
at a fixed value  of the 
energy $E_\nu$  depends both on  the  zenith and azimuth  
angles ($\theta_z$  and $\varphi$).
The angular  dependence originates  from two, or possibly
three sources:
\begin{enumerate}
\item The development of cosmic ray showers in the atmosphere.
\item Geomagnetic  effects  on the  primary cosmic ray flux.
\item Neutrino oscillations.
\end{enumerate}

The development of an hadronic  shower    induced by a primary particle
of  given  energy and mass  depends only on the zenith angle.
Cascades at large zenith angle develop in a relatively less 
dense part of the atmosphere, so that decay to neutrinos  is
enhanced at large angle.
%\footnote{In the GeV range, this effect
%is most important for products of muon decay.  The small geomagnetic effects   
%during the development  of  a shower  can be neglected here.}
In fact, the calculation of the secondary beam depends only 
on $|\cos\theta_z|$ because  a  line  of  sight 
entering the detector   from below the horizon  
with zenith angle $\theta_z^{out} >\pi/2$,
corresponds to   a  trajectory entering the atmosphere with 
$\theta_z^{in} = \pi -\theta_z^{out}$.
Apart from small  effects
due to the different average temperature profiles  of the atmosphere
at different  geographical locations,
the development of a shower does not depend on the  position 
of its impact point on the earth's surface. 
Thus, production of secondary particles in the atmosphere
is symmetric under the reflection $\cos\theta_z\leftrightarrow -\cos\theta_z$.

For a given energy spectrum, neutrino oscillations also depend
only on zenith angle; however, the dependence is strongly
asymmetric because the  pathlengths corresponding to 
the directions  $\pm\cos \theta_z$  are  very different.
For down--going particles, the neutrino pathlengths are in the range from
$\sim 10$ to $\sim 500$~km~\cite{pathlength}, whereas up--going
neutrinos have $L\sim 10^4$~km.

Geomagnetic effects modify the spectrum of primary cosmic
rays up to tens of GeV in a way that depends on azimuth as well as
zenith.  Since the neutrino flux is a convolution of the
primary spectrum with the yield of neutrinos per primary particle, neutrinos
with energies below a few GeV carry the imprint of these geomagnetic effects.
The geomagnetic field prevents  primary cosmic rays of low  rigidity
from reaching the atmosphere.  This suppression  depends  on both
the detector  location,
being lowest  (highest) at the magnetic poles (equator),
and on the line of sight  considered.
For  directions from  below the  horizon  the effect
must obviously be  calculated for the geomagnetic   field
at the position where the cosmic--ray  trajectories enter 
the volume of the atmosphere.   The nuclear  component  of cosmic
rays  is  positively  charged, and this introduces a
dependence on the azimuth angle,
the  celebrated  east--west effect. The  neutrino  flux is  highest
(lowest) for   directions  coming from the west (east). 

For most interactions of atmospheric neutrinos in present detectors
the direction of the neutrino is not fully reconstructed.
Typically, the direction of the charged lepton will be used to
indicate the direction of the event.  For detected leptons with momenta in
the interval $0.2 \le E_{\mu, e}  \le 1$~GeV
the east--west asymmetry   is of order  30\%,  after  taking into account
the dilution of the  effect   due to the broad distribution of
angles between the charged lepton and the neutrino.  An effect of
this size should be readily 
measurable  by high statistics experiments.
For  events  in the multi--GeV range,  when the initial
neutrino  energy  is of order of several GeV  the east-west effect
is reduced  to $\sim \;$ 10\%.
For a detailed study  it is necessary to  consider the exact  geographical 
location of the detector, and  the interval of 
neutrino  energy that is  detected.

In the presence of neutrino  oscillations  the zenith angle distribution
of the  detected events can be   significantly deformed; however,
the asymmetry in azimuth remains  unchanged to first order because
the neutrino pathlength does not depend on $\varphi$.  Only  to second
order would the deformation of the neutrino energy spectrum by oscillations
lead to a slight modification of the distribution in azimuth.
The azimuthal distribution depends only on
the filtering of the primary cosmic rays through the geomagnetic
field as viewed from each detector.
Study of the azimuthal dependence of neutrino interactions
can therefore  be a valuable 
diagnostic tool, both to validate calculations of the neutrino flux
and  to  check the quality of detector performance,
for example,  to demonstrate that the
determination of the  lepton directions  has the expected resolution.
Moreover,  since the geomagnetic  effects  
are the only known   mechanism   (besides  $\nu$--oscillations) that 
can produce an up--down  asymmetry  for the neutrino fluxes,
a measurement of   the east--west   effect   for neutrinos
would  establish  the size of the  geomagnetic  effects, 
and  greatly help in  limiting the possible importance of
geomagnetic  effects  on  the zenith angle distributions.
The existence of   distortions
of the zenith angle  distributions 
due to neutrino  oscillations  could then be more  clearly
identified.

This discussion can  be summarized  in the following equation:
\begin{eqnarray}
& & \phi_{\nu_\alpha}(E_\nu, \Omega_\nu,\vec{x}_d)  = 
\sum_A \phi_A (E_0)~ F_M [p_0/Z, \Omega_0, 
\vec{x}(\Omega_0)] \times  ~ \nonumber \\
& & ~~~~~~~~\sum_{\beta}
 {dn_{\nu_\beta} \over dE_\nu} (E_\nu, \; A, E_0, |\cos \theta_z|)  
~\langle P_{\nu_\beta \to \nu_\alpha} (E_\nu, \cos \theta_z, 
\{ m_j^2, U_{\alpha j} \}) \rangle 
\label{eq:nu-flux}
\end{eqnarray}
%In equation (\ref{eq:nu-flux})
where  $\phi_{\nu_\alpha}(E_\nu,\Omega_\nu, \vec{x}_d)$ is the flux of  
neutrinos  of  flavor $\alpha$  with energy 
$E_\nu$ and    direction $\Omega_\nu$  
observable in a detector  located at
a  position $\vec{x}_d$.
$\phi_A(E_0)$ is the flux of  primary cosmic  rays  of energy $E_0$
in the  vicinity of the  earth,  but at a distance  $r$ sufficiently  large
so that the effects of the geomagnetic  field  are  negligible 
($r \gtrsim 10\,r_\oplus$, where $r_\oplus$ is the radius of the earth). 
This  flux is isotropic  to  a very good approximation.
The cutoff factor $F_M$ takes into account the
effects  of the geomagnetic  field. It depends  on the rigidity 
$R = p_0/Z$ and  direction $\Omega_0$ 
of the  primary particle  and on the position 
$\vec{x}_{in}$   where its trajectory  first intersects
the atmosphere.
$dn_{\nu_\beta}/dE_\nu$ is the  average  number of
neutrinos  of flavor   $\beta$  produced   by  a primary particle
of mass $A$ and energy $E_0$. It  depends  on  the mass,
energy and  zenith angle  of the primary particle.  Finally,
$\langle P_{\nu_\beta \to \nu_\alpha}\rangle$  is the oscillation 
probability for the transition $\nu_\beta \to \nu_\alpha$ averaged  
over the position of  creation of the  neutrinos.  It  depends
on the  energy  and trajectory  of the neutrino  and on the 
oscillation  parameters; that  is, the mass eigenvalues
$m_j$ and the mixing matrix $U_{\alpha j}$  that relates mass 
and flavor eigenstates.

We first discuss the calculation of the probability of penetration
through the geomagnetic field.  Then in \S3 we describe the
Monte Carlo convolution expressed in Eq.~\ref{eq:nu-flux}.  Results
and discussion follow.

\section {Geomagnetic effects}

To a first approximation the effects of the field can be 
described simply by a cutoff rigidity $R_c(\vec{x}_{d}, \Omega)$, which is a
function of  the detector position $\vec{x}$  and the
direction $\Omega$, such  that all
rigidities smaller (larger) than $R_c$  are  forbidden (allowed);  that is:
\begin{equation}
F_M (R,   \Omega, \vec{x}_d)  = \Theta [ R - R_c (\theta_z,
 \varphi, \vec{x}_d) ],
\label{eq:Stormer0}
\end {equation}
where $\Theta (x) $ is the  Heavyside function ($\Theta(x) = 0$ for $x < 1$,
$\Theta(x) = 1$ for $ x \ge 1$).

In the case of a  dipolar  magnetic field that fills the  entire
space,  it is possible to  compute the  cutoff  rigidity exactly:
\begin {equation}
R_c = R_S (r, \lambda_M, \theta_z, \varphi) =  \left ( {M \over 2 r^2 }
\right )  ~ \left \{  { \cos^4 \lambda_M \over
[1 + (1 + \cos^3 \lambda_M \sin \theta_z \sin \varphi)^{1/2}]^2 } \right \},
\label{eq:Stormer}
\end{equation}
where $M$ is the magnetic dipole moment, $r$ (the  distance 
from the  dipole center)
and $\lambda_M$  (the  magnetic  latitude)  describe the detector  position,
$\theta_z$ is the zenith  angle  and $\varphi$ is an  azimuthal  angle,
with $\varphi= 0$ (${\pi\over 2}$)  indicating the 
north (west) direction.  For the earth
$M \simeq 8.1 \times 10^{25}$~Gauss~cm$^3$, which corresponds to a polar
magnetic  field  of 0.62 Gauss. The quantity $M/(2 r_\oplus^2)
\simeq 59.4$~GV corresponds to the rigidity of a particle in a circular orbit
of radius $r_\oplus$ in the earth's magnetic equatorial plane.

  St\/{o}rmer's formula (Eq. \ref{eq:Stormer})
 gives a good idea of the magnitude of the
 geomagnetic cutoffs, but it has limited accuracy because the geomagnetic
 field is only approximately an offset dipole. The formula also generally
 underestimates the cutoffs because it neglects the shadow of the Earth, i.e.
 allows the penetration of charged particles with trajectories that would
 have intersected the surface of the Earth. More exact calculations can
 be done using the backtracking technique~\cite{backtracking}
 and more realistic models of the geomagnetic field \cite{IGRF}.

 In the backtracking technique,   to establish  if a particle
 with  charge   $Z$  and  momentum $p$ 
 traveling  from interplanetary space 
 can  reach  a final point $\vec{x}$   close to  the surface of the earth
 arriving from  the direction $\Omega$, 
 one integrates the equation of motion for a particle with
 opposite charge  and  reflected  momentum   starting from this
 final  position.
 If the  backtracked  anti--particle  reaches  infinity, we can 
 assume that  the    trajectory is allowed ($F_M = 1$),
 if the backtracked particle is trapped in the geomagnetic field 
 or if its trajectory intersects
 the surface ($r =  r_\oplus$) the trajectory is considered forbidden
 ($F_M = 0$).
 Such a calculation was performed in Ref.~\cite{Roma1},   considering  as
 `trapped' those trajectories   that  remained  confined within 
  $r \le 30\ r_\oplus$    for a pathlength  longer $l \ge 500\ r_\oplus$.
 Instead of  the sharp  rigidity cutoff  predicted  by equation 
\ref{eq:Stormer},
 above which all particles from a particular direction reach the atmosphere,
 one ecounters  a quite  different  situation:
 in the vicinity of $R_S$ particle trajectories change rapidly with the
 rigidity, and the sharp cutoff is replaced with a series of allowed 
 ($F_M=1$) and  disallowed ($F_M=0$) rigidities -- the penumbra region.

  The function $F_M$ used in this paper is calculated by backtracking
 particles for a set of rigidities at $\Delta \cos{\theta_z}$ of 0.02 and
 $\Delta \phi$ of 5$^\circ$. The results are than averaged for angular
 bins of $\Delta \cos{\theta_z}$ = 0.1 and $\Delta \phi$ of 30$^\circ$,
 i.e. using the cutoffs for 36 directions for every rigidity value.
 The cutoffs are thus replaced with the probability per angular bin
 for a cosmic ray of given rigidity to reach vertical altitude of
 20 km and interact in the atmosphere.

\section {Montecarlo Calculation}

The calculation of the neutrino flux that we use here is described
in Refs.~\cite{Agrawal,Roma2}.  
Yields of neutrinos are calculated separately
for a grid of energies for primary protons and neutrons.
In the energy range of interest here, approximately 80\%
of the incident nucleons are free protons.  Most of the
rest are neutrons and protons in primary alpha particles.
For each direction (20 bins of $\cos\theta_z$ and 12 bins of $\phi$)
the yields are folded with the primary spectrum to obtain the
neutrino flux.  The primary spectrum is weighted with the cutoffs
averaged over the $\cos\theta_z$--$\phi$ bin.

We assume  that the neutrinos are collinear  with the  primary
cosmic ray particles that produce them.  The angle  $\theta_{0\nu}$
of the neutrino with respect to the shower axis,
for approximately half of the muon neutrinos,
can be  schematically  written as:
$\theta_{0\nu} = \theta_{\pi} \oplus \theta_{\pi\nu}$,
where $\theta_{\pi}$  is the angle 
  between the  parent meson (most of the time  
a  charged   pion) and the  primary particle, and $\theta_{\pi\nu}$
the angle between  the meson  and  the neutrino;
for electron neutrinos and for the other half of the muon neutrinos
one  has  to consider  a   two--decay chain, and  the angle  between
primary  particle and  neutrinos  is
$\theta_{0\nu} = \theta_{\pi} \oplus \theta_{\pi\mu} \oplus
 \theta_{\mu\nu}$.
Since the maximum  $p_\perp$ kinematically  allowed in  a $\pi^\pm$ 
($\mu^\pm$)  decay is 30  (50)~MeV, the dominant contribution
to  the neutrino  angle   comes from the transverse  momentum of the parent 
meson  (of  order 300~MeV):
\begin{equation}
\langle \theta_{0\nu} \rangle \simeq
\langle \theta_{\pi} \rangle \simeq 
{\langle p_{\perp,\pi} \rangle \over E_\pi} \simeq
{300~{\rm MeV} \over 4~E_\nu} \simeq {4.3^\circ \over E_\nu({\rm GeV}) }
\end {equation}
where we have  used the fact  that on  average the neutrino has
approximately one--quarter of the parent pion energy. 
The angle $\theta_{0\nu}$ is smaller that  the   angle 
$\theta_{\ell\nu}$ between  the detected  charged lepton  and the neutrino,
and its  neglect   does  not  introduce significant errors
in the predictions of the geomagnetic  effects  on the 
angular  distribution of the charged  leptons.

The next step is to treat the interaction of the neutrinos in
the detector and find the direction of the produced leptons.
For this purpose we use the quasi--elastic
and single pion neutrino cross sections as calculated
in Ref.~\cite{LLS} including corrections for 
nuclear target.  For deep inelastic scattering we use
the structure functions of Ref.~\cite{GRV}. Neutral
current interactions are neglected.  The direction
of each neutrino is chosen randomly within the bin and the
direction of the outgoing electron or muon is then chosen randomly using
the appropriate differential cross section.

To provide realistic and relevant examples of the angular dependence,
we consider two classes of events, applying 
cuts similar  to those of Super--Kamiokande (SK)~\cite{SuperK}.
As a low--energy sample, we use quasi--elastic simulated events
in which electrons (muons) have momenta in the interval
$0.1 < p_e < 1.33$~GeV/c  ($0.2 < p_\mu < 1.33$~GeV/c).
We compare this low energy sample 
to the `single--ring'  subset of the sub--GeV data of SK.
As a high--energy sample, we compare all events 
with lepton momenta in the interval $1.33<p_\ell<10$~GeV
with the multi--GeV data of SK.
Our definitions are of course not precisely equivalent to
the experimental classifications.  For example, 
the single--ring events in the data include some multi--particle
events in which only one is visible.  Conversely, some quasi--elastic events 
would be excluded from the single--ring sample because of
an energetic recoil proton.  We did check that both these contributions
are small and that they do not significantly alter the angular distributions
of leptons.  Thus we believe that the simplified cuts we make on the
Monte Carlo 
are adequate for our purposes to illustrate the expected
angular dependence of the two categories of leptons.  As a confirmation,
we can compare the number of events in our cuts with the
corresponding cuts in the SK data.
We find 689 sub--GeV electrons and 1050 sub--GeV muons (single ring only)
as compared to
789 and 1185 in 25.5 kT--years that the Super--Kamiokande Collaboration
report~\cite{SuperK} from their simulation using the same neutrino flux.
For the same exposure, we find a total of 1034 multi--GeV leptons
as compared to 1176 in the SK simulation.

 We show in Fig. 1a the distributions of neutrino energies that give
 rise to the two classes of events. The average neutrino energy for
 our sub--GeV muon sample os 0.8 GeV as compared to 5.7 GeV for
 the higher energy class.
The order--of--magnitude difference in energy corresponds to a similar
difference in $L/E$ and makes the atmospheric neutrino beam a
powerful probe of oscillations in an interval of parameter
space with large mixing and $10^{-3}\le\Delta m^2\le10^{-2}$~eV$^2$.
This follows from the well--known expression for survival of a neutrino
flavor in vacuum, which in a two--neutrino example is
\begin{equation}
P_{\nu_\beta \to \nu_\alpha}\;=\;1\,-\,\sin^22\theta\sin^2
\left [1.27{\Delta m^2(eV^2)\,L_{km}\over E_{GeV}}\right],
\label{osc}
\end{equation}
together with the large differences in pathlength between
up--going and down--going neutrinos.

\begin{figure}[!hbt]
\centerline{\psfig{figure=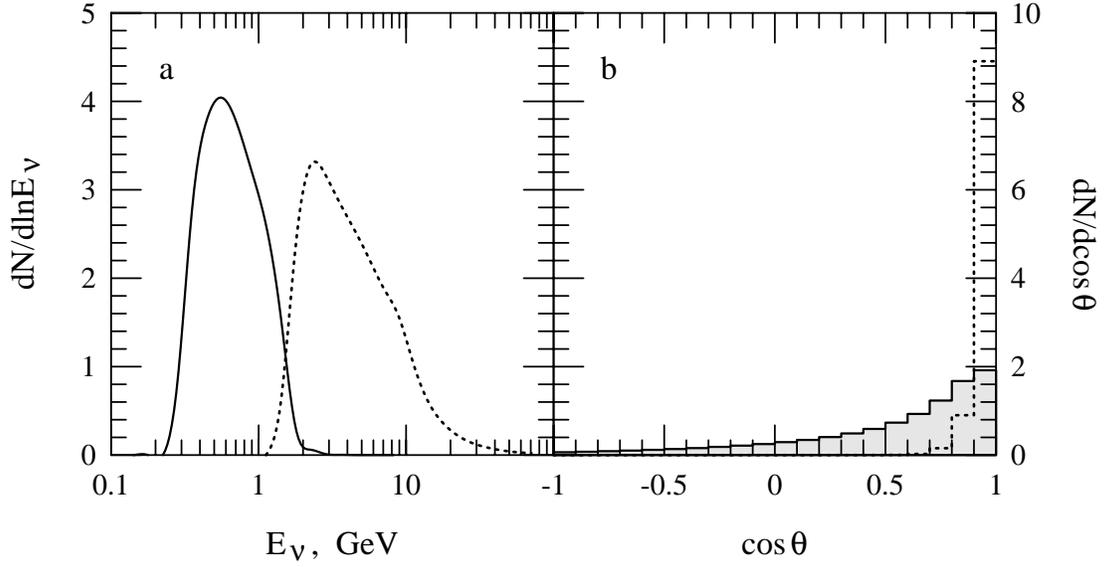,width=15cm}}
\caption{\em
  a). Distributions of neutrino energies that give rise to the sub--GeV
 (solid line)  and multi--GeV (dotted line) muon samples at Kamioka --
 see text for the definitions of the two groups of events in this calculation.
  b). Distribution of $\cos{\theta}$ ($\theta$ is the angle between the
 neutrino and muon direction) for the same two muon samples, same coding.
\label{obso95}}
\end{figure}

 In Fig.~\ref{obso95}b we show the distribution of the angle  between
 the detected  muon and the parent neutrino.
 The average values of $\cos{\theta}$ for the sub-- and multi--GeV
 samples are 0.53 and 0.97 respectively. 
 Normally only the charged lepton is detected, and because of the
 angle $\theta_{\ell \nu}$ with respect to the parent neutrino direction
 there is a smearing of the angular distribution of the neutrinos
 which is significant for the sub--GeV sample. 
 Only detectors with high granularity can measure the recoiling
 nucleons (or more complicated hadronic final states) and reconstruct
 the neutrino energy and direction. Such a measurement is potentially
 highly valuable in the search for neutrino oscillations. 
 
\begin{figure}[!hbt]
\centerline{\psfig{figure=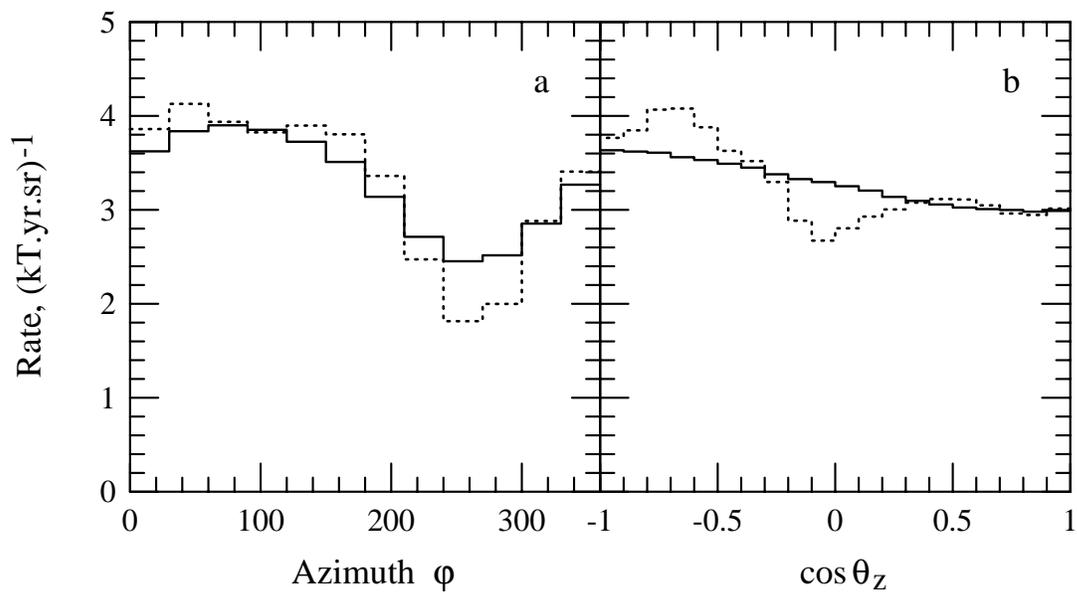,width=15cm}}
\caption{\em
  a) Azimuthal distributions for the sub--GeV muons (see text) at
 Kamioka (solid line)
 and for their parent atmospheric neutrinos (dotted line) averaged over
 the zenith angle $\theta_z$. b) $\cos\theta_z$ distributions  (averaged
 over the azimuth angle $\varphi$) for the
 same muon and neutrino samples. 
\label{obso91} }
\end{figure}

 We illustrate the effect of the angular smearing  in Fig.~\ref{obso91}
 by showing the azimuthal
and zenith angle distributions for sub--GeV muons (solid lines)
as compared to the same distributions for their parent neutrinos (dotted lines)
at the location of SK.  The azimuth is defined
so that $0^\circ$ corresponds to events from the north.  Azimuthal
angle increases counter--clockwise.  The most prominent feature
is the excess of events from the west.  The $\cos\theta_z$ dependence
for sub--GeV events at Kamioka is dominated by the high local geomagnetic
cutoffs.  We understand the slight excess of events from below 
($\cos\theta_z<0$) as
arising from the fact that the local geomagnetic cutoffs, which
affect the down--going events, are generally higher than the cutoffs
averaged over the opposite hemisphere of the earth that regulate
the up--going events.  We note that both geomagnetic poles are below
the horizon at Kamioka and therefore contribute to the relative
excess of events for negative $\cos\theta_z$.
The depression of the neutrino flux near
the horizon indicates that, for the sub--GeV events, 
the geomagnetic suppression more than compensates for the enhanced
production of neutrinos from muon decay in this same angular region.

In Fig. 3 we compare the azimuthal dependence for sub--GeV and
multi--GeV events for four intervals of $\cos\theta_z$ of equal solid angle.
The multi--GeV sample is sufficiently high in energy that the geomagnetic
effects are much reduced.

\section {Results and Discussion}

Geomagnetic location is of great importance for the nature of
the fluxes and angular distributions of low--energy events.
We illustrate this in Fig. 4 by comparing 
the angular distributions expected in the absence of oscillations
for sub--GeV muons at Kamioka
with that expected at Soudan~\cite{Soudan} or SNO~\cite{SNO}.  
The latter two experiments are near the north geomagnetic pole, so
the flux of down--going events is significantly higher than at Kamioka.
Moreover, the east--west effect is nearly absent for events coming from
above.  The sky--maps are in local coordinates with the local
zenith at the top and the local nadir direction at the bottom
of each map.

The angular--dependence of the neutrino events recently reported
from Super--Kamiokande~\cite{SuperK} has suggested several interpretations.
Among these, the simplest possibilities are $\nu_\mu\leftrightarrow\nu_\tau$
and  $\nu_\mu\leftrightarrow\nu_{sterile}$~\cite{ALL93,Liu-Smirnov,Foot-Volkas}.
In Fig. 5 we show the expected zenith angle dependence for sub--GeV
and multi--GeV neutrino induced muons at Kamioka (using our definition).  
The solid
line shows the result for no--oscillations.  In the multi--GeV
sample, the expected enhancement near the horizontal is clearly visible.
The three broken lines show the results expected according to Eq.~\ref{osc}
for full mixing assuming  $\nu_\mu\leftrightarrow\nu_\tau$ 
with $\Delta m^2 = 10^{-2},\,10^{-2.5},\,{\rm and}\, 10^{-3}$~eV$^2$.
 The distortion of the zenith angle distribution produced 
 by neutrino oscillations depends on the oscillation parameters.
 Vertical up--going muons have the same suppression
 $\sim \, 1 - {1 \over 2} \sin^2\,{2\theta}$ because
 of the averaging of oscillations on the long pathlength.
 The shape of the suppression factor as a function of zenith
 angle depends strongly on $\Delta m^2$ and is different
 in the two samples, reflecting the order of magnitude
 difference in the typical energy of the neutrinos that give
 rise to the events. These features are potentially distinguishable
 with the future high statistics data of Super--Kamiokande.

Whereas the shape of the zenith angle dependence strongly reflects
assumptions about oscillations, the azimuthal dependence at fixed zenith is
practically the same for all oscillations scenarios.  We show this in
Fig. 6 for the sub--GeV muons (Kamioka, our definition).  
The four panels are for regions 
of equal solid angle   of increasing zenith angle 
 from the vertically down--going quadrant to the vertically up--going
quadrant.  Typical pathlengths in the two down--going quadrants 
are $\sim30$~km and $\sim 300$~km, with large
variations due to neutrino--lepton scattering angle as well as
the relatively broad distributions of production height~\cite{pathlength}.
The up--going quadrants have pathlengths of order $10^4$~km, with an
admixture of shorter pathlengths near the horizontal direction.

Table 1 gives a quantitative summary of the east--west effect
for neutrino induced muons at Kamioka.  The west/east ratio is
$\approx 1.35$ for down--going and $\approx 1.28$ for up--going, sub--GeV
muons.  The ratio is $\approx 1.10$ for the multi--GeV muons.  
These ratios have a negligible dependence on the nature of the
assumed oscillation.  For this reason, study of the azimuthal
dependence of neutrino interactions should provide an important
probe of the systematics of searches for neutrino oscillations
with the atmospheric neutrino beam.  In addition, study of
the azimuthal and zenith angle dependence of electrons should
be a sensitive test of whether (as suggested in 
Refs.~\cite{Harrison-Perkins-Scott,Foot-Volkas})
oscillations on terrestrial scales also involve electron neutrinos
to some extent.

 {\bf Acknowledgements} The authors acknowledge helpful discussions with
 E.~Kearns. PL thanks BRI for its hospitality during the completion of
 this work. The research of TKG and TS is supported in part
 by the U.S. Department of Energy under Grant Number  DE~FG02~01ER~4062.

%\newpage
%%%%%%%%%%%%%%%%%%%%%%%%%%%%%%%%%%%%%%%%

\begin{table}[!htb]

\caption {Average muon rates (in units (Kt~yr~sr)$^{-1}$)
in four  solid  angle  quadrants.
The rates are calculated in the absence of oscillations  and 
for $\nu_\mu \to \nu_\tau$ oscillations with
maximal  mixing and three  values of $\Delta m^2$.
\label{tab-1} }

\medskip
\begin{tabular} {l | c c c c c | c c c c }
~~ & \multicolumn{4}{c}{sub--GeV} & ~ & \multicolumn{4}{c}{multi--GeV} \\
\tableline
$\Delta m^2$  (eV$^2$)~ & \multicolumn{2}{c}{downgoing} &
 \multicolumn{2}{c}{upgoing}  &  ~ & 
 \multicolumn{2}{c}{downgoing} & \multicolumn{2}{c}{upgoing}\\
~~ & west & east & west & east & ~ & west & east & west & east \\ 
\tableline
 No--osc.    & 3.56 & 2.59 & 3.92 & 3.06 & ~ & 2.55 & 2.31 & 2.55 & 2.33 \\
 10$^{-3}$   & 2.98 & 2.20 & 2.22 & 1.70 & ~ & 2.46 & 2.23 & 1.38 & 1.25 \\
 10$^{-2.5}$ & 2.67 & 2.00 & 2.08 & 1.62 & ~ & 2.35 & 2.15 & 1.31 & 1.20 \\
 10$^{-2}$   & 2.06 & 1.54 & 1.94 & 1.52 & ~ & 2.15 & 1.98 & 1.28 & 1.17 \\
\end{tabular}
\end{table}
%%%%%%%%%%%%%%%%%%%%%%%%%%%%%%%%%%%%%%%%%%

\newpage
%%%%%%%%%%%%%%%%%%%%%%%%%%%%%%%%%%%%%%%
%%%%%%%%%%%%%%%%%%%%%%%%%%%%%%%%%%%%%%%

\newpage
%%%%%%%%%%%%%%%%%%%%%%%%%%%%%%%%%%
%%%%%%%%%%%%%%%%%%%%%%%%%%%%%%%%%%%

\newpage
%%%%%%%%%%%%%%%%%%%%%%%%%%%%%%%%%%
\begin{figure}[!hbt]
\centerline{\psfig{figure=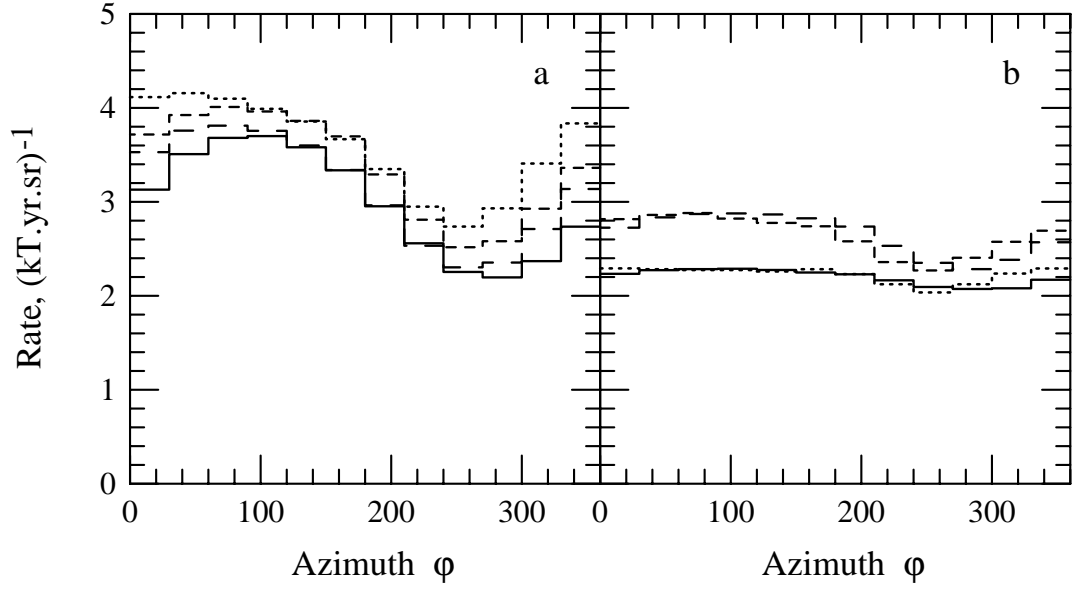,width=15cm}}
\caption{\em
  a) Azimuthal distributions for the sub--GeV muons (see text) at
 Kamioka for 
 four bins in $\cos{\theta_z}$: 1 to 0.5 (solid); 0.5 to 0. (dots);
 0. to --0.5 (short dash); --0.5 to --1. (long dash). 
  b) Azimuthal distributions for the multi--GeV muons at Kamioka,
 same coding.
\label{obso93}}
\end{figure}
%%%%%%%%%%%%%%%%%%%%%%%%%%%%%%%%%%%

\newpage
%%%%%%%%%%%%%%%%%%%%%%%%%%%%%%%%%%
\begin{figure}[!hbt]
\centerline{\psfig{figure=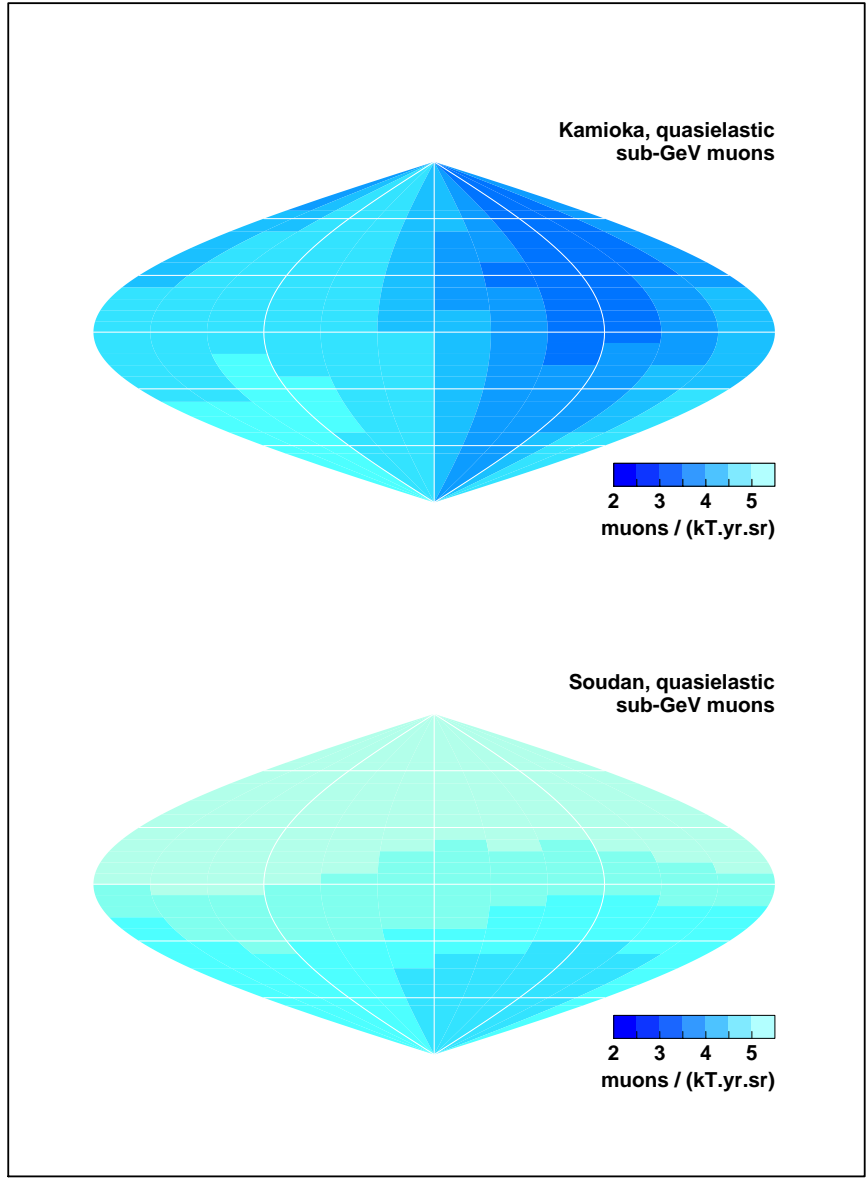,width=12cm}}
\caption{\em
 Two dimensional azimuthal and zenith angle distribution of the
 sub--GeV muons (see text) at Kamioka (top) and Soudan/SNO (bottom)
  in units of number of muons per kT.yr.sr. The top of the maps 
 corresponds to the local zenith and the bottom to the local nadir.
 The  north direction ($\varphi=0$)   corresponds to the
 edge of the map, south ($\varphi=180^\circ$)  to the vertical line 
 in the middle  with west ($\varphi$ = 90$^\circ$)  and east 
 ($\varphi= 270^\circ$)  to the left and right.
Notice  the smeared east--west effect in both maps.
\label{map}}
\end{figure}
%%%%%%%%%%%%%%%%%%%%%%%%%%%%%%%%%%%

\newpage
 %%%%%%%%%%%%%%%%%%%%%%%%%%%%%%%%%%
\begin{figure}[!hbt]
\centerline{\psfig{figure=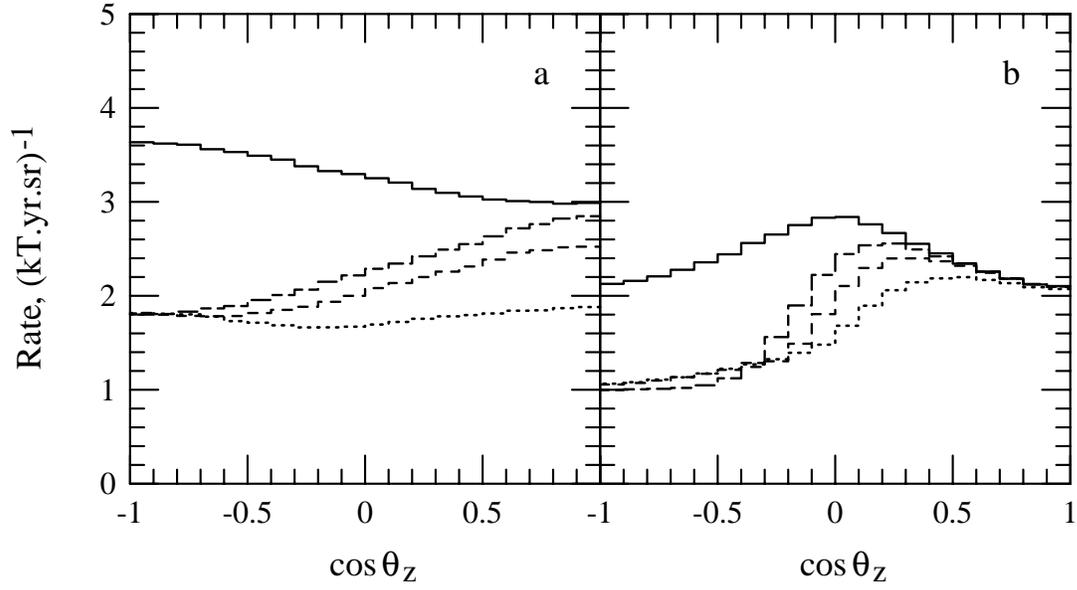,width=15cm}}
\caption{\em
  a) $\cos{\theta_z}$ distribution for the sub--GeV muons (see text) at
 Kamioka, averaged over azimuth -- solid line. The other histograms
 are for $\nu_\mu \rightarrow \nu_\tau$ oscillations with maximal mixing
 and $\Delta m^2$ = 10$^{-2}$ eV$^2$ (dots), 10$^{-2.5}$ eV$^2$
 (dashes) and 10$^{-3}$ eV$^2$ (dash--dash). 
  b) The same distributions for multi--GeV muons, same coding.
\label{obso94}}
\end{figure}
%%%%%%%%%%%%%%%%%%%%%%%%%%%%%%%%%%%

\newpage
%%%%%%%%%%%%%%%%%%%%%%%%%%%%%%%%%%
\begin{figure}[!hbt]
% \centerline{\psfig{figure=bartol$users:[rengel.tex.cross]pptot.ps,width=12cm}}
\centerline{\psfig{figure=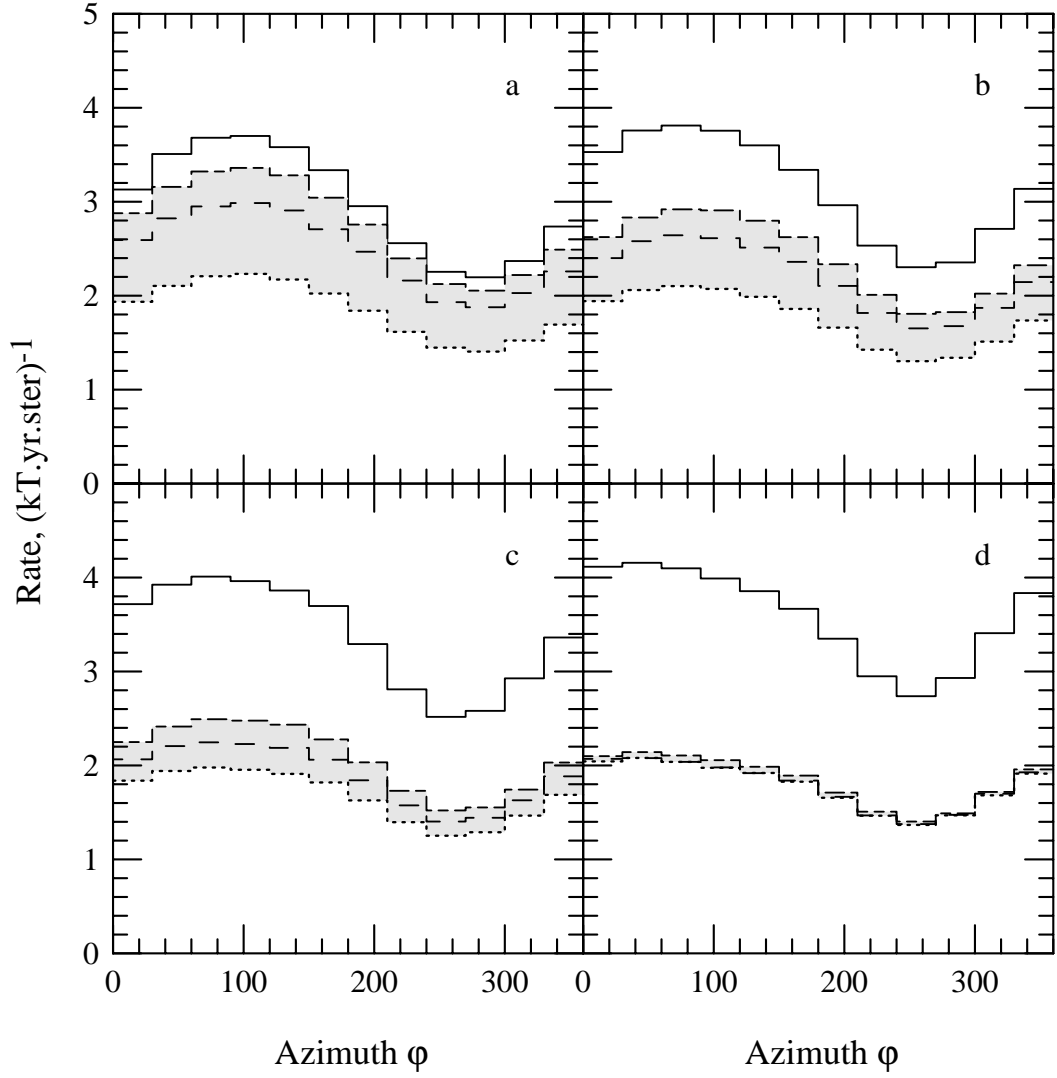,width=15cm}}
\caption{\em
  a) Azimuthal distributions for the sub--GeV muons (see text) at Kamioka
 for four bins in $\cos{\theta_z}$: 1 to 0.5 (a); 0.5 to 0. (b);
 0. to --0.5 (c); --0.5 to --1. (d) -- solid lines. The other histograms
 show the distributions in the presence of $\nu_\mu \rightarrow \nu_\tau$
 oscillations for maximal mixing and  and $\Delta m^2$ = 10$^{-2}$ eV$^2$ (dots),
 10$^{-2.5}$ eV$^2$ (dashes) and 10$^{-3}$ eV$^2$ (dash--dash). 
\label{obso92}}
\end{figure}
%%%%%%%%%%%%%%%%%%%%%%%%%%%%%%%%%%%

\end{document}